\documentclass{optica-article}

\journal{opticajournal} 

\articletype{Research Article}
\usepackage{orcidlink}
\usepackage{siunitx}
\usepackage{lineno}

\begin{document}

\title{Integrating high-precision and fringe-scale displacement sensing using heterodyne cavity-tracking}

\author{Shreevathsa Chalathadka Subrahmanya,\authormark{*}\orcidlink{0000-0002-9207-4669} Christian Darsow-Fromm,\orcidlink{0000-0001-9602-0388} and Oliver Gerberding\orcidlink{0000-0001-7740-2698}}
\address{Institute of Experimental Physics, University of Hamburg, Luruper Chaussee 149, 22761 Hamburg, Germany}
\email{\authormark{*}shreevathsa.subrahmanya@uni-hamburg.de}

\begin{abstract*} 
We present a heterodyne stabilized cavity-based interferometer scheme that can serve as a compact and high-sensitivity displacement sensor with a fringe-scale operating range.
The technique, in principle, can reach a sub-femtometer noise floor and an operating range on the order of one laser wavelength at $\lambda\approx$ \qty{1}{\micro\meter}.
With our current experimental setup, we achieve a sensitivity of about \qty{260}{fm/\sqrt{Hz}} at \qty{1}{\hertz} and \qty{46}{fm/\sqrt{Hz}} at around \qty{130}{\hertz}.
By probing a length actuated cavity, we demonstrate six orders of magnitude of dynamic range for displacement measurement, reaching a maximum motion of \qty{0.15}{\micro\meter}.
The tracking bandwidth and displacement range are limited by analog effects in the signal digitization and are extendable in the future. 
\end{abstract*}

\section{Introduction}
Displacement and inertial sensors are used to measure motion in many high-precision experiments. 
A prime example of such a use case is ground-based gravitational wave detectors like Advanced LIGO~\cite{aligo2015}, Advanced Virgo~\cite{Acernese2014}, KAGRA~\cite{Akutsu2020a}, GEO600~\cite{Lck2010}, and the future detectors like Einstein Telescope~\cite{Punturo2010} and Cosmic Explorer~\cite{Reitze2019,Hall2021}.
The seismic isolation systems of such detectors use various inertial and local displacement sensors~\cite{Matichard2015}.
They are used in their isolation and control systems to reduce the transmission of ground motion to the core optics of the interferometer, thereby keeping them `free-falling.'
Improving the sensitivity of these sensors is crucial to reducing associated noise sources in current and future detectors that limit the detector sensitivity at low frequencies. 
Laser interferometric sensors are prime candidates for realizing more sensitive displacement and, in turn, inertial measurements.
However, they need to be able to operate over a sufficient range to cope with the actual motions in the corresponding experiment. 
For any interferometer scheme, it is beneficial to have a high dynamic range, defined as the ratio between the maximum displacement that can be measured (or operating range) and the sensitivity (or precision). 
Finally, the interferometers must be compact to integrate with compact inertial sensors~\cite{Carter2024} and seismic isolation systems.

Several techniques have been investigated to realize compact interferometric sensing~\cite{Watchi2018}, for example, those based on strong laser frequency modulations that use simple optical setups at the cost of complex data processing~\cite{Gerberding2015,Eckhardt2024,Smetana2022}.
Homodyne quadrature-based interferometers are another approach and have also demonstrated high precision (with a sensitivity of about \qty{100}{fm/\sqrt{Hz}} at \qty{1}{\hertz}) and high-range or multi-fringe readout capabilities~\cite{Poar2011,Cooper2018}.
Heterodyne laser interferometers in compact setups with a sensitivity of \qty{11}{pm/\sqrt{Hz}} at \qty{0.1}{\hertz} are also reported~\cite{Zhang2022}.


We introduce and demonstrate a compact interferometric scheme suitable for high-precision displacement readouts, aiming for sub-femtometer noise levels at frequencies around \qty{1}{\hertz} and a fringe-scale (meaning, in the scale of a laser wavelength or an optical fringe) operating range.
This higher precision and moderate operating range are relevant for compact opto-mechanical sensors that operate in quiet environments or in-loop seismic isolation systems~\cite{Carter2024a,vanHeijningen2023}.
For example, an inertial noise of about \qty{0.1}{pm/\sqrt{Hz}} at \qty{50}{\hertz} will move the mechanical oscillator reported in~\cite{Carter2024} by roughly \qty{100}{\nano\meter} at its resonance (assuming a Q value of about \num{1e6}~\cite{Carter2024a}).
Such a moderate displacement range is trackable by our heterodyne cavity-tracking scheme.
A coating thermal noise limited displacement measurement with this oscillator, which is conceptually possible with our scheme, as described below, would be sufficient to realize a suspension thermal noise limited inertial sensor that is more sensitive than an L4C geophone at all frequencies between \qty{0.1}{\hertz} and \qty{100}{\hertz}.
We aim to bridge the gap between narrow-band and multi-fringe displacement sensing and achieve high precision in such applications.

Our scheme is based on the heterodyne laser frequency stabilization work by Eichholz \emph{et al.}~\cite{Eichholz2015}.
When laser interferometry is chosen for displacement sensing, the photon shot noise becomes the fundamental readout limitation for sensitivity~\cite{Lawall2000}.
Interferometric schemes that operate over one or multiple fringes are also limited by other readout noise, such as analog-to-digital converter (ADC) quantization noise~\cite{Eckhardt2022}, effectively limiting them to the femtometer level. 
In our scheme, the lasers are locked to optical cavities, which helps to reduce most of the associated readout noise contributions and shot noise to negligible levels.
Cavity locking techniques such as Pound-Drever-Hall (PDH)~\cite{Black2001} or heterodyne stabilization operate over a small displacement range, i.e., within the resonance of an optical cavity.
Here, we extend the overall operating range by letting the lasers follow the cavity motion and reading out the resulting large laser frequency changes.

The dynamic range for laser interferometric measurements has been pushed to extreme values with heterodyne laser interferometry, as developed for the Laser Interferometer Space Antenna (LISA)~\cite{Schwarze2019}. 
More than ten orders of magnitude of dynamic range were demonstrated by tracking the frequency change of a simulated, Doppler-shifted laser with a wavelength of \qty{1}{\micro\meter}, reaching a precision of \textless \qty{1}{pm/\sqrt{Hz}} with dynamics on the order of \qty{1}{\meter} and maximum speeds in the order of \qty{15}{\meter/\second}.

Our scheme uses the same readout method, a frequency-tracking phasemeter, but measures a frequency change that is not associated with a Doppler-shift. 
Rather, it is associated with a change in the free-spectral range of an optical cavity.
This enables us to maintain the dynamic range and shift it to smaller displacements and higher sensitivities.
The usage of optical cavities with a few centimeter lengths keeps this interferometric scheme compact and also enables its implementation as a readout of compact inertial sensors~\cite{Hines2023,Carter2024a,Carter2024}.

\section{Interferometric scheme}

The heterodyne cavity-tracking technique involves two lasers that are slightly frequency offset from one another.
Their interference forms a beat note at this offset frequency. This optical beat is sent to two optical cavities, with one of their end mirrors being the test mass we want to probe the displacement of.
The two laser frequencies are then controlled by locking each to the resonance of one of the two cavities.
If the length of one of the cavities changes, then the frequency that resonates inside that cavity also changes.
The control loop then makes the corresponding laser frequency follow this change.
Hence, in such an optical setup, monitoring the beat note gives us information on the relative length change of the optical cavities.

\begin{figure}[ht!]
\centering\includegraphics[width=10cm]{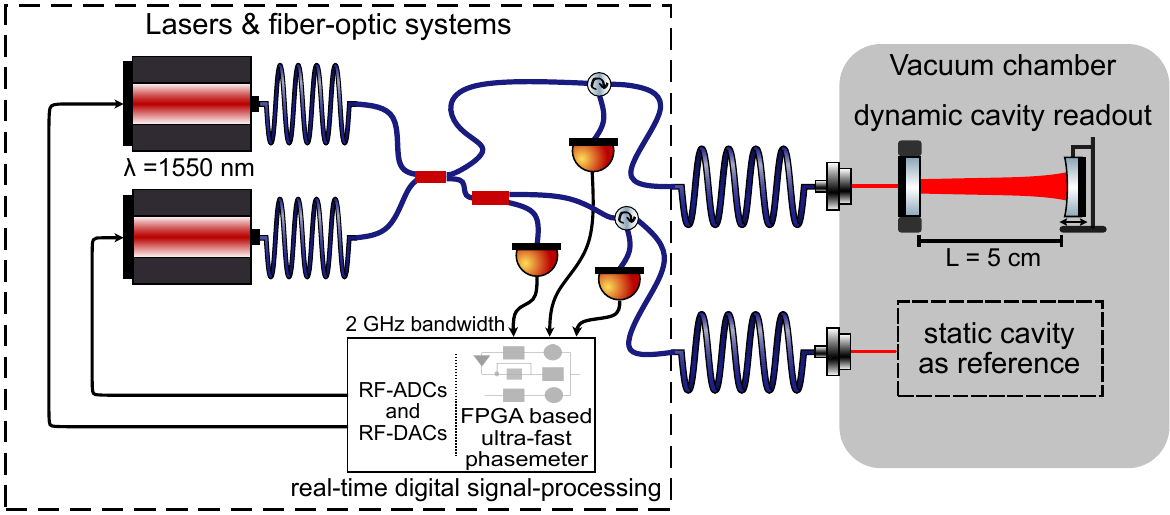}
\caption{Interferometric scheme of heterodyne cavity-tracking.
The beat note between the lasers and the signals reflected from the cavities are detected using photo-detectors and sent to the phasemeter.
One laser is locked to the resonance of a static cavity, acting as the reference. 
The other laser is made to follow the resonance of the dynamic cavity.
The beat note frequency is our signal of interest for displacement sensing.
(Modified from \cite{vanHeijningen2023}, with the permission of AIP Publishing.)}
\label{fig:ifo_scheme}
\end{figure}

Further, if we have one ultra-stable cavity in the setup as a reference, only the length fluctuation of the other dynamic cavity is encoded in the beat note.
This way, the interferometric scheme can be used for displacement sensing by dynamic tracking of the frequency change of the optical cavities. 
A simple and compact way of realizing such an experimental setup is sketched in Fig.\,(\ref{fig:ifo_scheme}).

In this case, the one-way motion of the end-mirror of the dynamic cavity is given by
\begin{equation}
\label{eq:1}
    \Delta L = \frac{\Delta f}{f} \cdot L
\end{equation}
where $\Delta f$ is the heterodyne beat note, $f$ is the absolute laser frequency, and $L$ is the absolute physical length of the dynamic cavity.
As we assume macroscopic cavity lengths (longer than a few \unit{\centi\meter}), we discard here corrections due to the comparatively small Gouy phase on Eq.\,(\ref{eq:1}) and assume that the microscopic mirror displacements have no relevant effects on the cavity mode or coupling.

\subsection{Heterodyne cavity-locking and readout}
There are numerous methods to keep the lasers locked to the resonance of an optical cavity.
The well-known and often-used method for this purpose is the PDH technique~\cite{Black2001}.
In the context of LISA, such studies are carried out using about \qty{20}{\centi\meter} long cavities~\cite{Jersey2024}.
Even though this PDH locking would be an option for our purpose, here we use the heterodyne cavity-locking technique~\cite{Eichholz2015} that integrates our readout and control architectures.
This stabilization scheme also mitigates the requirement of electro-optic modulators (EOMs), simplifying the experimental setup.
Contrary to the other interferometric methods for displacement sensing, we are not interested in the signal strength of the photo-detectors.
The information of primary interest is encoded in the beat note frequency.
The precision and dynamic range with which we can read out this frequency is significantly greater than if the displacement information were encoded in an optical power change, as is the case for many other interferometric displacement sensing schemes~\cite{Eckhardt2022}.
These other interferometer techniques are directly limited by the dynamic range of the ADCs. While frequency measurements are also limited by ADCs, this limit applies only to the phase resolution of the beat frequency and does not limit the frequency range (the latter is limited by the ADC sampling speed and analog bandwidth). 

Heterodyne locking uses the heterodyne beat note and the cavity reflection to generate the error signal for the control loop.
The latter signal is demodulated using the former one, which leaves the cavity interaction phase information as the discriminator.
Hence, to enable the locking of the lasers to the cavities, we need to track the heterodyne beat note between the lasers. 
From Eq.\,(\ref{eq:1}), it follows that the maximum displacement tracking range for a given optical setup can be expressed as
\begin{equation}
\label{eq:2}
    \Delta L_\mathrm{max} = \frac{\lambda}{2} \cdot \frac{\mathit{BW}}{\mathit{FSR}} = L \cdot \frac{\mathit{BW}}{f} 
\end{equation}
with $\lambda$ being the wavelength of the laser having a frequency $f$, $\mathit{BW}$ as the bandwidth of the frequency readout instrument, and $\mathit{FSR}$ as the free-spectral range of the dynamic cavity of physical length $L$.
For example, for an optical cavity length of \qty{5}{\centi\meter}, the beat note will shift by \qty{2}{\giga\hertz} when the cavity length is changed by $\lambda/3$.
To read out such a displacement from this interferometric scheme, one would need a frequency-tracking instrument with a bandwidth high enough to cope with the corresponding operating range.
The higher the readout bandwidth, the higher the dynamic tracking range.

For this purpose, we use the in-house developed GHz Phasemeter, a frequency- (or phase-) tracking instrument implemented on a Zynq UltraScale+ RFSoC ZCU111 evaluation kit~\cite{subrahmanya2024}. 
This readout system has \qty{2.048}{\giga\hertz} of signal bandwidth, \qty{2}{\mega\hertz} of tracking bandwidth, and an exceptionally large tracking speed demonstrated up to \qty{240}{\giga\hertz/\second}.
Hence, the GHz Phasemeter with such a high signal bandwidth makes it possible to read out a fringe-scale displacement range.
The high tracking bandwidth of the phasemeter allows the usage of diode lasers with a high non-stationary frequency noise spectrum for the readout.
The high tracking speed ensures the generation of error signals that keep the lasers locked to the cavities while following the rapid changes in the cavity length.

The readout noise floor of the GHz Phasemeter is comparable to that of the LISA phasemeter at measurement frequencies around \qty{1}{\hertz} for input signal frequencies around \qty{25}{\mega\hertz}, reaching a precision of \unit{\micro\hertz}~\cite{subrahmanya2024}. 
For signal frequencies at \qty{2}{\giga\hertz}, this precision is reduced for the GHz Phasemeter to about \qty{100}{\micro\hertz}.
The signal bandwidth of the LISA phasemeter, being \qty{25}{\mega\hertz}, gives a maximum dynamic range on the order of about (\num{25e6})\,$/$\,(\num{1e-6}) $=$ \num{2.5e13}.
The GHz Phasemeter has a similar maximum dynamic range of (\num{2.048e9})\,$/$\,(\num{1e-4}) $=$ \num{2.048e13} for the readout (assuming the worst-case precision for the GHz Phasemeter currently demonstrated).

\section{Digital heterodyne controller}

\begin{figure}[ht!]
\centering\includegraphics[width=10cm]{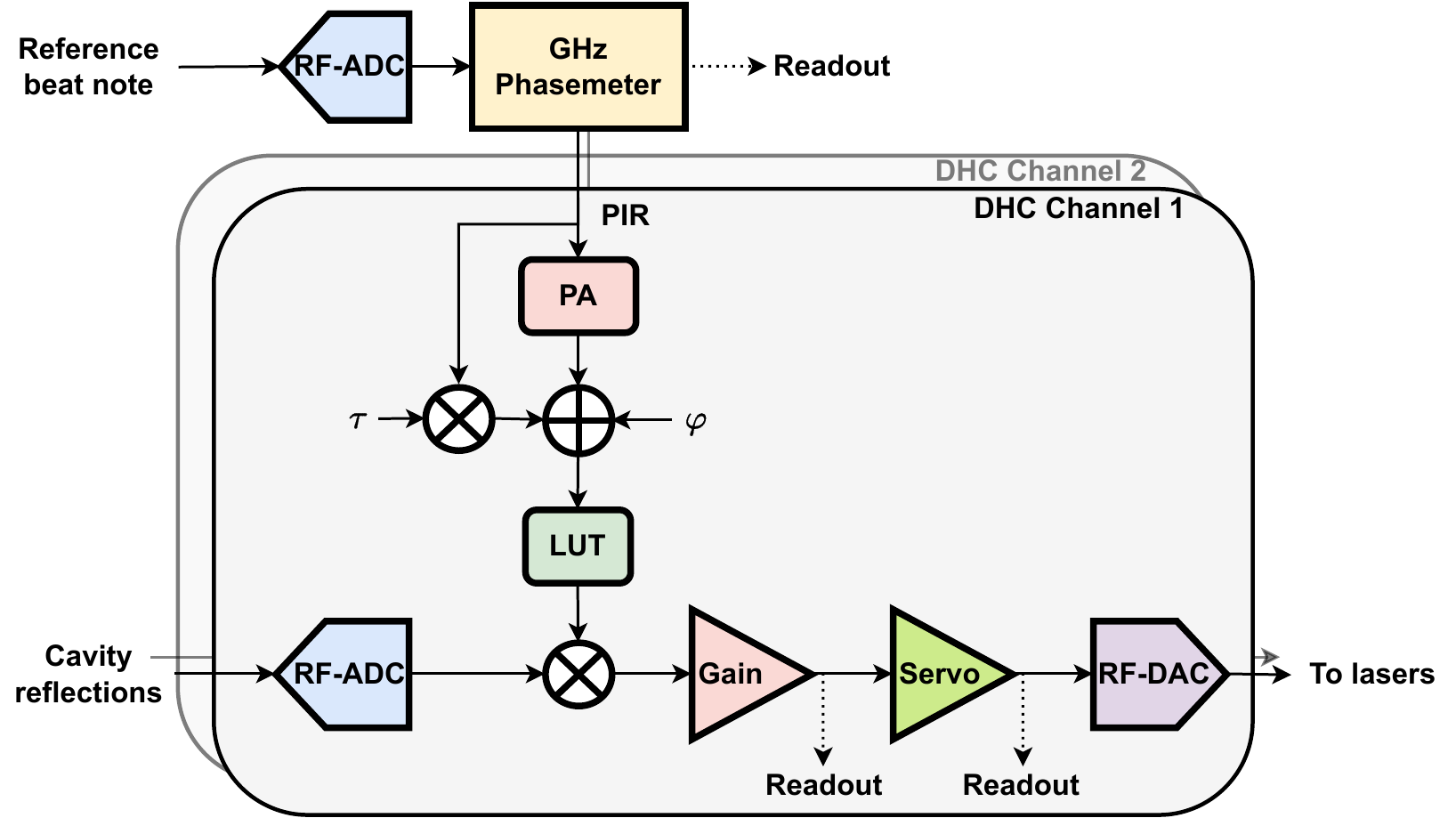}
\caption{Block diagram of the digital heterodyne controller. Reference beat note and reflections from the cavities are inputs to RF-ADCs. 
The actuation signals for the lasers come out of RF-DACs. Markers for data readout are shown. 
PIR: phase increment register, PA: phase accumulator, LUT: look-up table, $\tau$: frequency-dependent demodulation phase correction term, $\varphi$: frequency-independent demodulation phase correction term.}
\label{fig:DHC}
\end{figure}

As mentioned before, our solution integrates readout and control architectures.
Following this, we have implemented the laser-locking control loop in the programmable logic (PL) of the same RFSoC using a field programmable gate array (FPGA) algorithm. 
We call it the digital heterodyne controller (DHC).
The block diagram of DHC is sketched in Fig.\,(\ref{fig:DHC}).

The beat note between the lasers is coupled to a radio-frequency analog-to-digital converter (RF-ADC) on the evaluation board.
This tone is then tracked by the GHz Phasemeter, our readout system.
The current frequency value of the phase-locked loop (PLL), stored in the phase increment register (PIR), serves as the reference for the heterodyne demodulation.
In theory, for such demodulation to work, it would be enough to directly demodulate the signal reflected from the cavity with the reference beat note. 
However, in practical setups, the path length difference between the two signals getting mixed causes an additional unwanted slow oscillation on top of the actual error signal.
Both optical and electrical path length differences contribute to this effect.
For example, assume a \qty{1}{\meter} path length difference between cavity reflection and reference beat note purely due to electrical cable length.
The arrival time delay between these two signals at the mixer would then be about \qty{5}{\nano\second}.
If we then assume a fast scan of laser frequency covering a significant part of the FSR, say \qty{100}{\mega\hertz/\second}, then it results in a \qty{0.5}{\hertz} oscillation in the error signal because of the delay.

From the above example, it is clear that there should not be any path length mismatch, which is very difficult to realize in experimental setups.
To address this issue, we have implemented a frequency-dependent delay compensation in the FPGA algorithm. 
Instead of using the reference beat note directly as the demodulation signal, the PIR is used as the starting point. 
It is then incremented (or accumulated) to generate a phase value, forming the phase accumulator (PA) stage of DHC.
Before feeding this PA output to the look-up table (LUT) to create a demodulation signal, we implemented the feature to add external offsets to phase values.
The frequency-dependent correction technique takes an external value $\tau$ (corresponding to time delay) and multiplies it with the current frequency (given by PIR).
The product is then used as the correction term for the PA output.
Such a correction technique was proposed in~\cite{Eichholz2015}. Here we implemented it with the linear trend and demonstrated it over a large bandwidth.
The frequency-independent phase correction term $\varphi$ helps in further fine-tuning the correct demodulation phase. 
Higher order corrections beyond linear frequency dependence have not yet been implemented but are possible in principle.

The \qty{4.096}{GSPS} (giga samples per second) RF-ADCs are configured to give \num{8} data samples in parallel at \qty{512}{\mega\hertz} which is the maximum possible processing speed.
To handle these eight signals simultaneously, the PA, LUT, and the mixer shown in Fig.\,(\ref{fig:DHC}) are eight processing stages in parallel, each operating in the PL at \qty{512}{\mega\hertz} (based on the architecture described in~\cite{subrahmanya2024}).
The outputs of the mixers should be low-pass filtered to get the error signal for the servo (Proportional-Integral controller).
The filtering involves a rolling average of \num{16} samples, implemented as consecutive additions, forming a finite impulse response filter. 
The servo output is passed to a radio-frequency digital-to-analog converter (RF-DAC) on the evaluation board, configured at \qty{0.512}{GSPS}.
Both RF-ADC and RF-DAC are differential in nature.
For RF-DAC differential outputs, we have a conventional Operational Amplifier-based in-house circuitry to get a single-ended signal for the laser controller. 
In the case of RF-ADC, to convert a single-ended photo-detector output into differential signals, we use the TRF1208 evaluation module from Texas Instruments.
To match with the RF-ADC input impedance of \qty{100}{\ohm}, we exchanged both the output resistors of the module to \qty{49.9}{\ohm}.
Nevertheless, the magnitude of the RF-ADC transfer function showed additional ripples on top of the general drop to higher frequencies, indicating the RF back-reflections due to impedance mismatch.
This currently limits our ability to match the demodulation phase and to keep constant control gains for large frequency changes.
This unwanted analog front-end effect could potentially degrade the overall performance.

To facilitate heterodyne cavity-locking for two laser systems, two DHC channels are running on RFSoC.  
The PL is controlled, and the data readout is done in real-time via Yocto Linux running on the programmable system part of RFSoC.
This enables communication from the external world via a Python interface for setting up the proper delay and phase offset values, boosting the error signal, or adjusting the servo gains.

\section{Experimental results}
\subsection{Heterodyne cavity-locking and readout}
Two continuously tunable external cavity diode lasers (CTL-1550 from Toptica Photonics AG), operating at \qty{1550}{\nano\meter} wavelength, were locked to two static optical cavities using DHC.
The lasers themselves are very sensitive to mechanical and acoustic noises and have a comparably high laser frequency noise~\cite{subrahmanya2024}.
Therefore, the laser heads sit on an active vibration isolation platform and are acoustically isolated using a foam enclosure.
The laser output is fiber-coupled, and we also use a fiber-optic coupler/splitter (PN1550R5A2 from Thorlabs) that allows fiber input to split into two outputs or vice versa to generate the beat note between the lasers.  All of the fibers and fiber optics use polarisation maintaining single-mode fibers with angled physical contact (APC) connectors.
Part of the beat note is picked off and detected by a high-speed fiber-coupled InGaAs detector (DET08CFC/M from Thorlabs) with a \qty{5}{\giga\hertz} bandwidth.
Similar photo-detectors are used to detect cavity reflections as well.

For the static cavities, we use high-reflective mirrors mounted in commercial kinematic mounts (POLARIS-K1E from Thorlabs) to form impedance-matched cavities with a finesse of about \num{312}.
We chose a flat-curved configuration, with the end mirrors having a radius of curvature of \qty{0.3}{\meter}.
One of the static cavities has a macroscopic length of \qty{10}{\centi\meter}, while the other has \qty{5}{\centi\meter}.
A fringe visibility of about \qty{95}{\%} was achieved for both cavities. 
Higher order modes excited due to minor misalignment have sufficient spacing to the fundamental mode we lock to in this configuration.
Hence, we can discard their influence on the former~\cite{Boyd2024}.

Fiber feed-throughs bring the beat note between the lasers to the vacuum chamber.
Fiber-optic circulators (CIR1550PM-APC from Thorlabs), which are non-reciprocating, unidirectional, three-port devices, behind the feed-throughs help us in distinguishing between the laser input and the reflected light from the cavities.
The vacuum chamber hosts an aluminum enclosure acting as a thermal shield. 
Inside this thermal shield, we place the aligned cavities, along with beam-steering mirrors and mode-matching lenses.
A moderate vacuum level of about \qty{1.8e-2}{mbar} was maintained inside the chamber.
Both the piezo voltage and diode current of the lasers were actuated using the laser controllers.

\begin{figure}[ht!]
\centering\includegraphics[width=12cm]{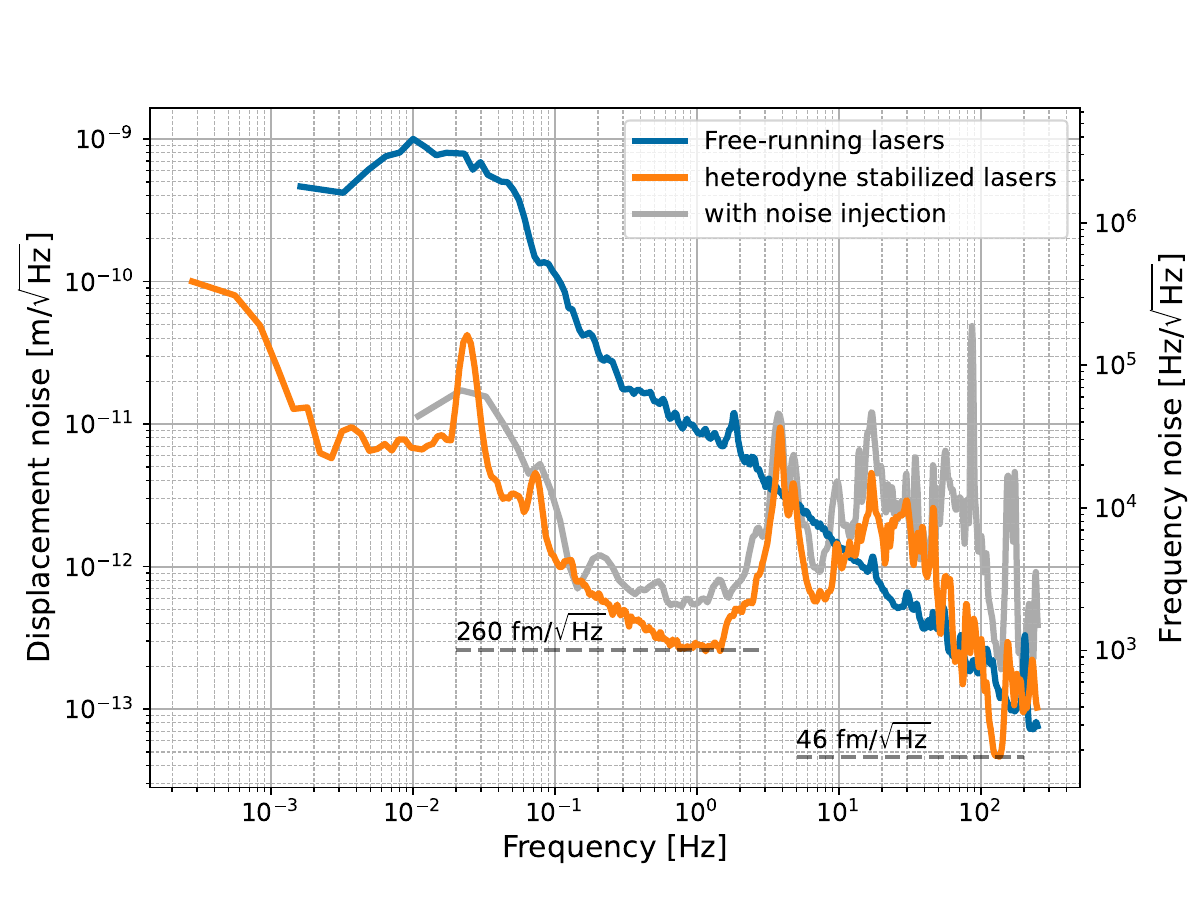}
\caption{Two continuously tunable lasers are stabilized to two optical cavities constructed using kinematic mirror mounts.
The \qty{10}{\centi\meter} cavity is taken as the reference, and the displacement noise of the \qty{5}{\centi\meter} cavity is plotted. 
Maximum sensitivity of \qty{46}{fm/\sqrt{Hz}} at around \qty{130}{\hertz} was achieved in the current setup.
For noise injection, a sinusoidal audio tone at \qty{86.1}{\hertz} was played in the proximity of the vacuum chamber.} 
\label{fig:HCL}
\end{figure}

Fig.\,(\ref{fig:HCL}) shows the experimental results when lasers are stabilized to the static cavities using the heterodyne locking technique.
Between \qty{0.1}{\hertz} and \qty{3}{\hertz}, the sensitivity is in the sub-picometer level.
The open-loop gain of each laser control loop was tuned to have a unity gain frequency (UGF) of around \qty{2}{\kilo\hertz}. 
The lasers stayed in lock for several hours, with no indication that longer locks were not possible.
Increasing the UGF to \qty{6}{\kilo\hertz} did not improve the sensitivity, confirming that we are not limited by the gain of the control loop.
Higher UGFs could not be achieved due to the limited control bandwidth available when using the laser controllers.  

In principle, this technique can achieve sub-femtometer level performance at frequencies above \qty{1}{\hertz} since contributions from shot noise are strongly suppressed by the optical cavity and contributions from ADC noise and electronic noise are negligible because of the chosen readout.
One can separate the noise discussion into two parts. The first is the noise associated with locking the lasers to the cavities. The second is the noise associated with the frequency readout of the laser beat note.
A detailed discussion of several noise sources that appear in laser locking loops is carried out in~\cite{Lam2010} for digital PDH locking and demonstrates a noise floor below \qty{1}{\hertz/\sqrt{Hz}}.
We expect a very similar noise coupling also for our stabilization technique implementation.
For our experimental setup, a readout precision of the heterodyne beat note of \qty{10}{\hertz} corresponds to a displacement readout noise contribution of the phasemeter of $< 10^{-15}$\,\unit{\meter}.
The ADC noise from the DHC channels and the shot noise of the photo-detectors that receive reflected light from the cavities still contribute to the overall noise budget.
With the chosen cavity parameters and input laser power levels of about \qty{10}{\milli\watt}, the displacement noise induced by the shot noise is in $10^{-17}$\,\unit{m/\sqrt{Hz}} regime.
This fundamental noise floor can be pushed down by increasing the laser power and using higher finesse cavities.
Length or phase noise in our fiber setup will also be present in the differential frequency measurement but couples weakly, especially at low frequencies. It can be well suppressed in compact setups. If an external ultra-stable laser is provided via fiber, then a corresponding fiber length stabilization can be utilized. We also assume there are no relevant phase dynamics in our fiber setup between the two laser frequencies. 
Hence, ideally, this technique can reveal the actual optical path length variation of the corresponding optical cavities to sub-femtometer precision.
Using stable cavities and/or more stable frequency references is necessary to demonstrate a lower achievable noise floor, and such future studies will have to verify that no additional, unexpected readout noise couplings arise in our technique in comparison to, e.g., PDH.




So far, we have not extensively attributed the observed noise to specific contributions.
Environmental thermal drifts and thermal noise from the mirror coatings likely take part at low and medium frequencies.
The prominent peak at about \qty{25}{\milli\hertz} could be attributed to the air conditioning system of our laboratory.
The fiber-optic components in the experimental setup introduce parasitic beams that interfere with the main interference, effectively resulting in parasitic etalons.
They produce rather large frequency instability, limiting the length stability at low frequencies~\cite{Shen2015}.
This was confirmed by switching the fiber circulators with fiber couplers/splitters, which resulted in higher sensitivity at frequencies below \qty{0.1}{\hertz}.
Cavities with higher finesse values will reduce this effect in the future by increasing the signal-to-noise ratio of the nominal relative to the parasitic signals.
Some other potential noise sources are residual amplitude modulation fluctuations of the laser source and thermal noise (Brownian motion) in the cavities used.
These are not rigorously analyzed here but descriptions in the context of PDH locking are available in the literature~\cite{Numata2004,Amairi2013,Kryuchkov2020}. 
At the high-frequency end, the mechanical instability of our cavities likely dominate the measurements, with a clear indication of acoustic noises influencing the stability of the cavity length as well. An acoustic noise injection test confirmed this claim (see Fig.\,(\ref{fig:HCL})).

\subsection{Heterodyne cavity-tracking}
\begin{figure}[ht!]
\centering\includegraphics[width=8cm]{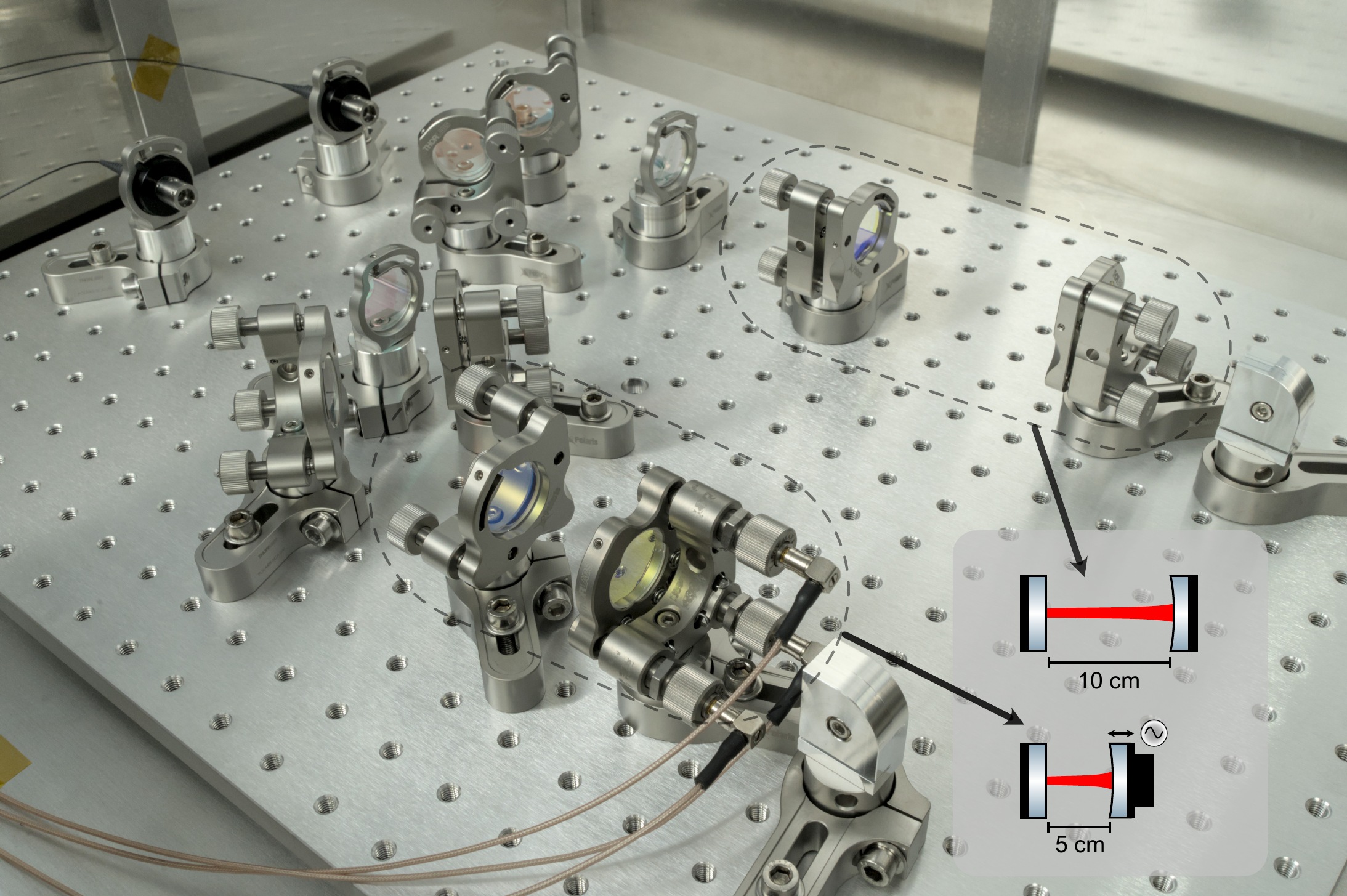}
\caption{Experimental setup for the heterodyne cavity-tracking. The beat note between the lasers is coupled via optical fibers. 
The aligned setup is placed inside a thermal shield, which goes into the vacuum chamber.}
\label{fig:lab}
\end{figure}

After verifying the heterodyne cavity-locking technique and understanding the current displacement readout noise floor, 
we moved to experimentally demonstrate the heterodyne cavity-tracking and probe the achievable operating range by using our GHz Phasemeter and DHC.
The end-mirror mount of the \qty{5}{\centi\meter} cavity is now replaced with a mount having three piezoelectric adjusters (POLARIS-K1S3P from Thorlabs).
A photograph of the experimental setup is shown in Fig.\,(\ref{fig:lab}).

\begin{figure}[ht!]
\centering\includegraphics[width=10cm]{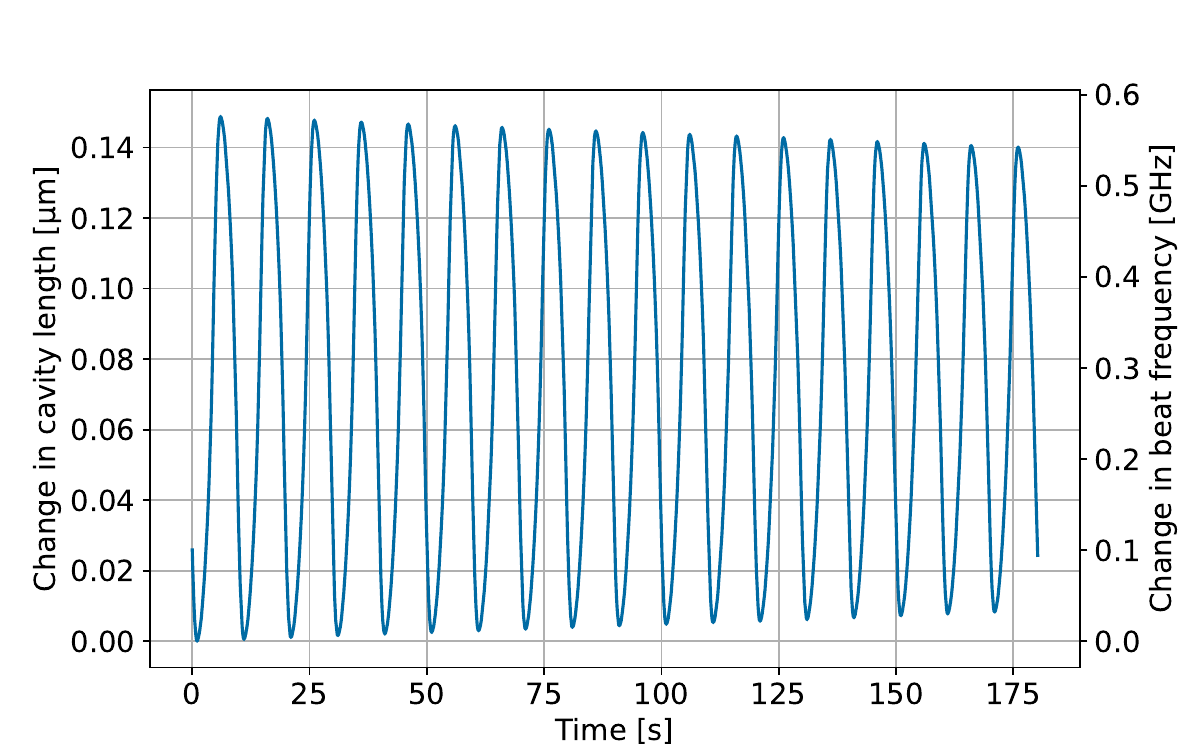}
\caption{Time series data of the beat note between the lasers when the cavity length was modulated using piezoelectric elements.
The corresponding length change of the cavity is later calculated and is given for reference.
An external sinusoidal scan signal at \qty{0.1}{\hertz} with \qty{0.5}{\volt} amplitude is given to the piezo controller (MDT693B from Thorlabs), which then controls all three piezoelectric adjusters.}
\label{fig:HCT}
\end{figure}

Once the lasers are locked to the cavities, a sinusoidal signal is applied to all three piezoelectric adjusters.
This, in turn, modulates the cavity length, and the corresponding laser follows this motion.
By tracking the beat note, we can then use Eq.\,(\ref{eq:1}) to get the displacement of the dynamic cavity end mirror with respect to the static cavity length.
Time series data of one such measurement is shown in Fig.\,(\ref{fig:HCT}).
The deviation from the sinusoidal shape is most likely because of the non-uniform expansion and compression of piezo elements to each other, resulting in additional tilt effects that create an effective, slow modulation of the cavity length modulation, visible here as decreasing modulation depth.
The before-mentioned analog front-end issue of RF-ADCs can also influence this.

The laser-locking was proven to work over a wide range of beat note frequencies, from some hundreds of \qty{}{\mega\hertz} to close to \qty{2}{\giga\hertz}.
For convenience and to show the dynamic tracking, we kept the beat note at around \qty{500}{\mega\hertz}.
When the cavity mirror moves by an order of a tenth of a micron, the beat note gets shifted by a fraction of a GHz.
The high-bandwidth readout and control system is capable of keeping the lasers in lock and, hence, acting as a displacement sensor over an operating range of about \qty{0.15}{\micro\meter}.
Considering the minimum noise floor achieved with the static cavities, this demonstrates a dynamic range between range and sensitivity of about six orders of magnitude.
By using newer RFSoCs with higher bandwidth ADCs or by implementing techniques like ADC frequency interleaving, one can increase the detection bandwidth and, hence, the operating range.

The demonstrated displacement range is limited by the error signal optimization in terms of time delay compensation and uncorrected effects in the analog signal digitization chain.
In our experiment, we could remove most of the oscillations in the error signals due to time delay, but not completely.
We suspect to be currently limited by the step sizes of the tuning parameters we implemented in the PL, requiring higher bit resolution for finer tuning in the future.
Also, the general drop of the ADC transfer function magnitude to higher frequencies and its phase response variation (ripple) are not taken into account and corrected in our current implementation.
This will limit the wide-band tuning and, hence, the operating range.
It is also important to have enough laser-locking bandwidth for high dynamic range applications.
Increasing this bandwidth is possible by introducing additional actuators like EOMs. 
From the RFSoC side, it would be possible to reach UGFs of some hundreds of \qty{}{\kilo\hertz}, assuming the control loop bandwidth would be limited by the delays of the RF-ADCs and RF-DACs, but this is not yet tested.
The jitter of the phasemeter clock and differential phase noise between the reference beat and the cavity reflection that does not result from the laser-cavity interaction are potential noise sources that can also affect displacement sensing~\cite{Eichholz2015}.
The use of impedance-matched cavities and more stable reference clocks can help reduce these influences should we find them limiting in the future.

\section{Conclusion}
We have demonstrated a heterodyne stabilized cavity-based interferometric scheme as a high-precision and fringe-scale displacement sensor.
This scheme can improve the readout of compact inertial sensors currently studied and in development.
The fiber-coupled setup, along with readout and control integrated into an RFSoC evaluation board, make our system compact and easily portable. 
While our scheme requires one laser per optical cavity being read out, we assume the high bandwidth of our readout and laser locking loop enables the usage of relatively noisy and, therefore, comparably inexpensive lasers. 
To use this as a readout for other inertial sensors, one would merely need to construct a cavity with one mirror attached to the proof mass of the sensor.

The demonstrated readout noise floor of about \qty{260}{fm/\sqrt{Hz}} at \qty{1}{\hertz} and \qty{46}{fm/\sqrt{Hz}} at around \qty{130}{\hertz} looks very promising as the initial results for the current experimental setup.
The maximum precision we reached here is comparable to the exceptionally low residual displacement measurement noise of \qty{32.1}{fm/\sqrt{Hz}} achieved down to much lower frequencies by the LISA Pathfinder mission~\cite{Armano2021}.
With a dynamic cavity, we have shown the dynamic tracking capability over a fringe-scale range and identified both the analog front-end and the measurement bandwidth of our ADCs as current limitations.
Improvements in the handling of the analog front-end effects and future RFSoCs with even higher bandwidth indicate that a full interferometric fringe operating range can be achievable in the near future for cavities as short as \qty{5}{\centi\meter}.
By using stable reference cavities and better thermal and mechanical isolations, one can further improve the displacement readout sensitivity and work towards demonstrating sub-femtometer noise levels.
The influence of parasitic beams can be attenuated by using a free-beam setup or increasing the cavity finesse. 

\begin{backmatter}
\bmsection{Funding}
Deutsches Zentrum f\"ur Luft- und Raumfahrt (50OQ2001, 50OQ2302); Deutsche Forschungsgemeinschaft (390833306).

\bmsection{Acknowledgments}
The authors thank Marcel Beck for the helpful discussions during the course of this work, DESY-MSK group for their FPGA infrastructure, and Sina K\"ohlenbeck for the suggestions via the LIGO P\&P review process.
This work was supported in part by the Deutsche Forschungsgemeinschaft (DFG, German Research Foundation) under Germany’s Excellence Strategy—EXC 2121 ``Quantum Universe''—390833306 
and in part by the Deutsches Zentrum f\"ur Luft- und Raumfahrt (DLR) with funding from the Bundesministerium f\"ur Wirtschaft und Klimaschutz under project references 50OQ2001 and 50OQ2302.
The authors acknowledge financial support from the Open Access Publication Fund of Universit\"at Hamburg.

\bmsection{Disclosures}
The authors declare no conflicts of interest.

\bmsection{Data availability} 
Data underlying the results presented in this paper are not publicly available at this time but may be obtained from the authors upon reasonable request.

\end{backmatter}

\bibliography{ref}

\end{document}